\documentstyle[multicol,eqsecnum,aps]{revtex}

\renewcommand{\narrowtext}{\begin{multicols}{2}
\global\columnwidth20.5pc\noindent}
\renewcommand{\widetext}{\end{multicols}
\global\columnwidth42.5pc}
\begin{document}
\draft
\preprint{12 November 1997}
\title{Quantum Fluctuations to Cause the Breakdown of the Spin-$1$
       Haldane Phase}
\author{Shoji Yamamoto}
\address{Department of Physics, Faculty of Science, Okayama University,\\
         Tsushima, Okayama 700, Japan}
\date{Received \hspace{4cm}}
\maketitle
\begin{abstract}
   Investigating quantum fluctuations in the ground states of $S=1$
quantum antiferromagnetic spin chains described by the
bilinear-biquadratic Hamiltonian
${\cal H}=\sum_i
   \left[
     \mbox{\boldmath$S$}_{i} \cdot \mbox{\boldmath$S$}_{i+1} 
   +\beta
    (\mbox{\boldmath$S$}_{i} \cdot \mbox{\boldmath$S$}_{i+1})^2
   \right]$,
we study a mechanism of the breakdown of the Haldane phase.
Based on the valence-bond-solid structure, but replacing two links of
them by triplet bonds ({\it crackions}), we construct a trial wave
function which is singlet and translationally invariant, where the
crackion-crackion distance is regarded as a variational parameter.
At $\beta<1/3$, the minimization of the variational energy results in
a bound state of crackions, while at $\beta>1/3$, crackions come to be
set free from their bound state with increase of $\beta$ and the chain
length.
We point out that the breakdown of the Haldane phase with $\beta$
approaching $1$ can be attributed to the collapse of the bound state
and the growth of a short-range repulsive interaction between
crackions.
\end{abstract}

\narrowtext
\section{Introduction}\label{S:I}

   Since Haldane \cite{Hald1} proposed that the one-dimensional
spin-$S$ Heisenberg antiferromagnet should exhibit qualitatively
different properties according to whether $S$ is integer or half odd
integer, low-temperature properties of integer-spin chains have been
of great interest.
Various numerical tools \cite{Whit1,Sore1,Goli1,Scho1,Yama1} brought us
precise estimates of the excitation gap immediately above the ground
state and visualized the Haldane massive phase.
On the other hand, a rigorous treatment \cite{Affl1} performed by
Affleck, Kennedy, Lieb, and Tasaki (AKLT) considerably contributed to
understanding of the physical mechanism of this phenomenon.
They considered a special bilinear-biquadratic Hamiltonian of $S=1$,
\begin{equation}
   {\cal H}=\sum_i
      \left[
        \mbox{\boldmath$S$}_{i} \cdot \mbox{\boldmath$S$}_{i+1} 
      +\frac{1}{3}
       (\mbox{\boldmath$S$}_{i} \cdot \mbox{\boldmath$S$}_{i+1})^2
      \right] \,,
   \label{E:AKLTH}
\end{equation}
and constructed its ground state using valence bonds, which they call
a valence-bond-solid (VBS) state.
This model possesses the unique disordered ground state with a finite
gap to the excited states and therefore exhibits the typical nature
peculiar to the Haldane phase.
Keeping in mind that within a naive variational treatment
\cite{Nomu1,Kenn1,Schd1,Tots1,Yama2} the ground states of a certain
class of Hamiltonians are all approximated by the VBS state, we are
convinced that the appearance of the Haldane phase is generally
described in terms of the VBS picture.

   The AKLT model reveals that there lies a hidden topological order
\cite{Girv1,Tasa1} in the Haldane phase, which can be measured by
the string order parameter of den Nijs and Rommelse \cite{Nijs1},
\begin{equation}
   O_{\rm string}^z
      =\lim_{|i-j|\rightarrow\infty}
       \langle
       S_i^z
       \prod_{k=i}^{j-1}\exp[{\rm i}\pi S_k^z]
       S_j^z
       \rangle\,,
   \label{E:string}
\end{equation}
where $\langle\ \ \rangle$ denotes the expectation value in the ground
state.
Since the string order parameter takes its full value $4/9$ in the VBS
state, any excitation of the VBS state should more or less reduce the
hidden antiferromagnetic order.
A local defect in the VBS state is obtained by replacing its arbitrary
link by a triplet bond \cite{Knab1}.
This spin-$1$ excitation, which is called a {\it crackion}, has a
solitonic nature.
Actually, a moving crackion and a moving domain wall in the hidden
antiferromagnetic order both result in the same dispersion relation
\cite{Arov1,Fath1}.
Furthermore, Scharf and Mikeska \cite{Schf1} reported that modified
crackions, which they call dressed solitonic excitations, almost
perfectly reproduce the low-lying excitations of the AKLT model
\cite{Fath1}.
Not only variational treatments \cite{Gome1,Kohl1,Neug1} but also
numerical investigations \cite{Whit1,Taka1,Yama3,Taka2,Yama4,Scho2}
showed that the idea of domain walls appearing in the hidden order
well applies to the low-lying excitations of the Heisenberg
Hamiltonian as well.

   Thus the low-energy structure of $S=1$ antiferromagnetic spin chains
is well described by the VBS ground state and the solitonic excitations
in its hidden topological order.
However, we have, on the other hand, to recognize that once the
Hamiltonian deviates from the AKLT point, quantum fluctuations begin to
destroy the perfect hidden order.
At the Heisenberg point, for example, $O_{\rm string}^z$ is reduced to
$0.374$ \cite{Whit1,Yama2,Scho2}.
Although the VBS model elucidates the essential mechanism of the
appearance of the Haldane phase, the breakdown of the Haldane phase
is never understood without consideration of the quantum fluctuations.
Since a density-matrix renormalization-group calculation \cite{Scho2}
suggests that $O_{\rm string}^z$ persists as far as the system lies in the
Haldane phase, the breakdown mechanism must be revealed in clarifying how
the string order is reduced and disappears.
Why is $O_{\rm string}^z$ able to remain finite in spite of fluctuations?
What is the driving force of the collapse of the string order?
In answering these questions, we have first to recognize that there exist
two mechanisms to destroy the string order.
One may be called the thermal mechanism and explains how the string order
vanishes at finite temperatures.
The other may generically be called the quantum mechanism and explains how
the ground-state phase transitions occur.
The qualitative difference between them was pointed out by the present
author and Miyashita \cite{Yama3}.
Making use of quantum Monte Carlo snapshots, they demonstrated that
thermal fluctuations can cause the collapse of the string order, whereas
quantum fluctuations no more than reduce the string order
(Fig. \ref{F:illust1}).
It is worth mentioning that thermal and quantum fluctuations should be
distinguished only in integer-spin chains.
Figures \ref{F:illust1}(c) and \ref{F:illust1}(d) show that for $S=1/2$
both thermal and quantum fluctuations merely result in a pair of domain
walls in the antiferromagnetic order and therefore there is no
qualitative difference between them.
Thus we are well convinced that thermal fluctuations induced in the
$S=1$ Haldane phase indeed lead to the collapse of the string order.
On the other hand, quantum Monte Carlo observations \cite{Yama3} of the
ground state of the Heisenberg Hamiltonian suggest that there exists an
attractive interaction between topological defects, which confines
quantum fluctuations to producing local effects on the hidden order.
Furthermore the present author \cite{Yama5} analytically demonstrated
that quantum-machanically induced solitonic excitations are stabilized
into a strongly bound state at the Heisenberg point.
That is why the string order parameter persists, in spite of
fluctuations, in the ground state of the $S=1$ Heisenberg model.

   Now we have to recognize that the quantum breakdown of the Haldane
phase has remained unexplained.
How does the bound state of the solitonic excitations behave as the
Hamiltonian approaches critical points?
This is the interest in this article.
Because of two reasons, we take the $S=1$ bilinear-biquadratic
Hamiltonian
\begin{equation}
   {\cal H}=\sum_{i=1}^L
      \left[
        \mbox{\boldmath$S$}_{i} \cdot \mbox{\boldmath$S$}_{i+1} 
      +\beta
       (\mbox{\boldmath$S$}_{i} \cdot \mbox{\boldmath$S$}_{i+1})^2
      \right] \,,
   \label{E:H}
\end{equation}
on a ring of $L$ sites for our subject.
First, the model contains the AKLT point, which allows us to make a
variational approach at the idea of crackions appearing in the VBS
background.
Second, the model contains explicit critical points, which were
revealed by the Bethe-ansatz method.
One is the Uimin-Lai-Sutherland (ULS) point \cite{Uimi1,Lai1,Suth1} of
$\beta=1$ and the the other is Takhtajan-Babujian (TB) point
\cite{Takh1,Babu1} of $\beta=-1$.
Hence we can take advantage of the knowledge on the boundaries of
the Haldane phase in investigating the mechanism of its breakdown.
It is the growth of quantum fluctuations with $\beta$ moving from
$1/3$ to $1$ that we explicitly discuss here.
Another boundary of the Haldane phase, $\beta=-1$, seems to be so far
from the AKLT point as not to allow us to approach it with the same
scenario as demonstrated in this article.
We will briefly mention another possible scenario to destroy the
string order in the final section.

   Although quantum Monte Carlo snapshots are quite helpful in getting
a qualitative view of fluctuations \cite{Yama3}, they are not naively
available in the region of $\beta>0$ due to the negative-sign problem.
Therefore, in order to investigate quantum fluctuations, we make an
analytic approach constructing a physically-motivated trial wave
function for the ground state.
Here, we take little interest in obtaining a superior variational bound
\cite{Schd1} on the ground-state energy but lay a great emphasis on
clarifying how the pair-crackion fluctuations grow as $\beta$ moves
away from $1/3$.
The trial wave function is naive but interestingly suggests a probable
mechanism for the breakdown of the Haldane phase caused by quantum
fluctuations.

\section{Trial Wave Function}\label{S:TWF}

   Before constructing a trial wave function, we briefly review our
knowledge on the ground state of the model as a function of $\beta$.
For recent years, the model has vigorously been argued and up to
now turned out to exhibit at least three different phases:
\begin{description}
   \begin{description}
   \item[$-1<\beta <1${\rm :}]
      Haldane phase \cite{Hald1,Affl1} with a unique disordered ground
      state and a gapped spectrum.
   \item[$\beta< -1${\rm :}]
      Dimerized phase \cite{Soly1,Chub1,Xian1} with twofold
      degenerate ground states and a gapped spectrum.
   \item[$1<\beta${\rm :}]
      Trimerized phase \cite{Nomu1,Fath2,Xian2,Reed1} with threefold
      degenerate ground states.
      Whether the spectrum is gapped or gapless has less been settled
      so far.
   \end{description}
\end{description}
A qualitative description of the phase diagram is obtained by a simple
 variational treatment \cite{Nomu1,Schd1,Xian2}
employing three different balence-bond states,
\begin{mathletters}
   \begin{eqnarray}
      |{\rm VBS}(L)\rangle
      &=&{\rm Tr}
         \left[
         g_1^{s}\otimes g_2^{s}
         \otimes\cdots\otimes
         g_L^{s}
         \right]\,, \\
      |{\rm Dimer}(L)\rangle
      &=&{\rm Tr}
         \left[
         f_1^{s}\otimes f_2^{s}
         \otimes\cdots\otimes
         f_L^{s}
         \right]\,, \\
      |{\rm Trimer}(L)\rangle
      &=&{\rm Tr}
         \left[
         h_1^{s}\otimes h_2^{s}
         \otimes\cdots\otimes
         h_L^{s}
         \right]\,,
   \end{eqnarray}
   \label{E:VBSpre}
\end{mathletters}
$\!\!\!$where
\begin{mathletters}
   \begin{eqnarray}
      &&
      g_{i}^{s}=
      \left[
      \begin{array}{cc}
       -        |0\rangle_{i}
      &-\sqrt{2}|+\rangle_{i} \\
        \sqrt{2}|-\rangle_{i}
      &         |0\rangle_{i}
      \end{array}
      \right]\,, \\
      &&
      f_{2i-1}^{s}=
      \left[
      \begin{array}{ccc}
        |-\rangle_{2i-1}
      & |0\rangle_{2i-1}
      & |+\rangle_{2i-1}
      \end{array}
      \right]\,,
      \nonumber \\
      &&
      f_{2i}^{s}=
      {\rm T}\left[
      \begin{array}{ccc}
        |+\rangle_{2i}
      &-|0\rangle_{2i}
      & |-\rangle_{2i}
      \end{array}
      \right]\,,
      \\
      &&
      h_{3i-2}^{s}=
      \left[
      \begin{array}{ccc}
        \sqrt{2}|+\rangle_{3i-2} &
                |0\rangle_{3i-2} &
        \sqrt{2}|-\rangle_{3i-2}
      \end{array}
      \right]\,,
      \nonumber \\
      &&
      h_{3i-1}^{s}=
      \left[
      \begin{array}{ccc}
        0
      & \sqrt{2}|-\rangle_{3i-1}
      &-        |0\rangle_{3i-1} \\
       -\sqrt{2}|-\rangle_{3i-1}
      & 0
      & \sqrt{2}|+\rangle_{3i-1} \\
                |0\rangle_{3i-1}
      &-\sqrt{2}|+\rangle_{3i-1}
      & 0
      \end{array}
      \right]\,,
      \nonumber \\
      &&
      h_{3i}^{s}=
      {\rm T}\left[
      \begin{array}{ccc}
        \sqrt{2}|+\rangle_{3i} &
                |0\rangle_{3i} &
        \sqrt{2}|-\rangle_{3i}
      \end{array}
      \right]\,,
   \end{eqnarray}
   \label{E:gpre}
\end{mathletters}
$\!\!\!\!$with $|+\rangle_i$, $|0\rangle_i$, $|-\rangle_i$ being the
$S_i^z$ eigenstates for eigenvalues $1$, $0$, $-1$, respectively.
$|{\rm VBS}(L)\rangle$ is the VBS state of AKLT type, which is
homogeneous.
$|{\rm Dimer}(L)\rangle$ and $|{\rm Trimer}(L)\rangle$ represent
balence-bond states which are dimerized and trimerized, respectively.
The linear combination of these three states,
\widetext
\begin{equation}
   |\Phi(L;\theta,\phi)\rangle
      =\cos\theta\,
       \frac{|{\rm VBS}(L)\rangle}
            {|\!|{\rm VBS}(L)|\!|}
      +\sin\theta\,\cos\phi\,
       \frac{|{\rm Dimer}(L)\rangle}
            {|\!|{\rm Dimer}(L)|\!|}
      +\sin\theta\,\sin\phi\,
       \frac{|{\rm Trimer}(L)\rangle}
            {|\!|{\rm Trimer}(L)|\!|} \,,
   \label{E:Trialpre}
\end{equation}
may be a naive but suggestive variational wave function for the ground
state of the present Hamiltonian, where $|\!| A |\!|$ denotes the norm
of the state vector $|A\rangle$.
Because of the asymptotic orthogonality between the different
balence-bond states, $|\Phi(L;\theta,\phi)\rangle$ is correctly
normalized in the thermodynamic limit.
The variational energy is obtained as
\begin{equation}
   \lim_{L\rightarrow\infty}
   \frac{\langle\Phi(L;\theta,\phi)
         |{\cal H}|
         \Phi(L;\theta,\phi)\rangle}{L}
      =-\frac{4}{3}+2\beta+\sin^2\theta
       \left[
        \frac{1}{3}+\frac{2}{3}\beta
       +\sin^2\phi
        \left(
        \frac{1}{3}-\frac{14}{9}\beta
        \right)
       \right]\,,
   \label{E:VEpre}
\end{equation}
\narrowtext
which leads to a simple solution,
\begin{equation}
   \left.
   \begin{array}{ll}
      \theta=\frac{\displaystyle\pi}{\displaystyle 2}\,,\ 
      \phi=0
    & \ (\beta<-\frac{\displaystyle 1}{\displaystyle 2})\,, \\
      \theta=0
    & \ (-\frac{\displaystyle 1}{\displaystyle 2}
       <\beta
       <\frac{\displaystyle 3}{\displaystyle 4})\,, \\
      \theta=\frac{\displaystyle\pi}{\displaystyle 2}\,,\ 
      \phi=\frac{\displaystyle\pi}{\displaystyle 2}
    & \ (\frac{\displaystyle 3}{\displaystyle 4}<\beta)\,.
   \end{array}
   \right.
   \label{E:solpre}
\end{equation}
Thus we expect that in a certain region around the AKLT point, the
low-energy physics of the model may be described in terms of the VBS
picture.
Although the model encounters a commensurate-incommensurate crossover
\cite{Scho2,Burs1} at a certain positive value of $\beta$, the gapped
spectrum and the string order parameter both persist in the whole
region between the TB ($\beta=-1$) and the ULS ($\beta=1$) points
\cite{Scho2}.

   We propose an idea of recognizing the breakdown of the Haldane
phase with $\beta$ moving away from the AKLT point as the growth of
quantum fluctuations in the hidden antiferromagnetic order.
The crackion with its spin projection $\lambda$ ($\lambda=+$, $0$, $-$)
is created at the bond between sites $i$ and $i+1$ replacing $g_i^s$
by $g_i^\lambda$ \cite{Tots1}, where
\widetext
\begin{equation}
   g^{+}=
   \left[
   \begin{array}{cc}
      \sqrt{2}|+\rangle & 0 \\
          -|0\rangle    & 0
   \end{array}
   \right]\,,\ \
   g^{0}=
   \left[
   \begin{array}{cc}
          -|0\rangle    &  \sqrt{2}|+\rangle \\
      \sqrt{2}|-\rangle &     -|0\rangle
   \end{array}
   \right]\,,\ \
   g^{-}=
   \left[
   \begin{array}{cc}
       0 &    -|0\rangle     \\
       0 & \sqrt{2}|-\rangle
   \end{array}
   \right]\,.
   \label{E:gtrip}
\end{equation}
The site indices have been omitted in Eq. (\ref{E:gtrip}) for the sake
of simplicity.
In order to describe the ground state, we introduce a trial wave
function of zero momentum with a pair of crackions,
\begin{equation}
   |\Psi(L;l,\alpha)\rangle
      =\frac{1}{\sqrt{L}}\sum_{i=1}^L
       \left[
          \frac
          {|\Psi_{i,i+l}^{+-}(L)\rangle}
          {|\!|\Psi_{i,i+l}^{+-}(L)|\!|}
         +\frac
          {|\Psi_{i,i+l}^{-+}(L)\rangle}
          {|\!|\Psi_{i,i+l}^{-+}(L)|\!|}
         -\alpha
          \frac
          {|\Psi_{i,i+l}^{00}(L)\rangle}
          {|\!|\Psi_{i,i+l}^{00}(L)|\!|}
       \right] \,,
   \label{E:VWF}
\end{equation}
\narrowtext
where
\begin{equation}
   |\Psi_{i,i+l}^{\lambda\mu}(L)\rangle =
   \left\{
   \begin{array}{l}
      {\rm Tr}
      \left[
      g_1^s\otimes\cdots\otimes g_{i-1}^s\otimes
      g_{i}^{\lambda}\otimes g_{i+1}^s\otimes\cdots
      \right.\\
      \ \ \ 
      \left.
      \otimes g_{i+l-1}^s\otimes g_{i+l}^{\mu}
      \otimes g_{i+l+1}^s\otimes\cdots\otimes g_{L}^s
      \right]\\
      \qquad\qquad\qquad\qquad\quad
      {\rm for}\ 1\leq l\leq L/2\,,\\
      {\rm Tr}
      \left[
      g_1^s\otimes\cdots\otimes g_{L}^s
      \right]
      \qquad\qquad\ \; 
      {\rm for}\ l=0\,.
   \end{array}
   \right.
   \label{E:2Cstate}
\end{equation}
Here, the crackion-crackion distance $l$ is regarded as a variational
parameter, whereas $\alpha$ has just been introduced so as to
symmetrize the two crackions into a spin singlet.
The condition of the total spin singlet,
\begin{equation}
   \sum_{i,j=1}^L
   \langle\Psi(L;l,\alpha)
   |\mbox{\boldmath$S$}_{i}\cdot\mbox{\boldmath$S$}_{j}|
   \Psi(L;l,\alpha)\rangle
   =0 \,,
   \label{E:constS}
\end{equation}
results in
\begin{equation}
   \alpha=\alpha_0
      \equiv
       \sqrt{
       \frac
       {3^L+3(-1)^L}
       {3^L+(-1)^l 3^{L-l}+(-1)^{L-l} 3^l+(-1)^L}} \,.
   \label{E:alpha0}
\end{equation}
We note that $\alpha_0$ generally deviates from unity.
This is because a crackion has its internal structure and spreads
over neighboring sites.
It is convincing that $\alpha_0$ coincides with unity only in the
limit of $L\rightarrow\infty$ and $l(\leq L/2)\rightarrow\infty$.
On the other hand, the minimization of the variational energy with
respect to $\alpha$,
\begin{equation}
   \frac{\partial}{\partial\alpha}
   \frac
   {\langle\Psi(L;l,\alpha)
    |{\cal H}|
    \Psi(L;l,\alpha)\rangle}
   {\langle\Psi(L;l,\alpha)|
    \Psi(L;l,\alpha)\rangle}
   =0 \,,
   \label{E:constE}
\end{equation}
also results in Eq. (\ref{E:alpha0}).
In this sense the trial wave function employed is reasonable enough
to investigate the ground-state fluctuations of the present
Hamiltonian.

   Hereafter we fix $\alpha$ to $\alpha_0$ and calculate the
variational energy as a function of $L$, $l$, and $\beta$,
\begin{equation}
   E(L;l,\beta)
      \equiv
      \frac
      {\langle\Psi(L;l,\alpha_0)|
       {\cal H}
       |\Psi(L;l,\alpha_0)\rangle}
      {\langle\Psi(L;l,\alpha_0)
       |\Psi(L;l,\alpha_0)\rangle}\,.
   \label{E:valE}
\end{equation}
The stabilization of the two-crackion state is measured by the energy
difference
\begin{equation}
   {\mit\Delta}E(L;l,\beta)
      \equiv E(L;l,\beta)-E(L;0,\beta)\,.
   \label{E:stab}
\end{equation}
The variational energy (\ref{E:valE}) is constructed from the quantum
averages
$\langle\Psi_{i,i+l}^{\lambda\mu}(L)
 |\mbox{\boldmath$S$}_{m}\cdot\mbox{\boldmath$S$}_{m+1}|
 \Psi_{j,j+l}^{\kappa\nu}(L)\rangle$
and
$\langle\Psi_{i,i+l}^{\lambda\mu}(L)
 |(\mbox{\boldmath$S$}_{m}\cdot\mbox{\boldmath$S$}_{m+1})^2|
 \Psi_{j,j+l}^{\kappa\nu}(L)\rangle$,
which are represented in the form of matrix-product type and
therefore calculated by the use of the transfer-matrix technique
\cite{Yama5,Tots2,Klum1,Kole1}.
The calculation is straightforward but exhausting due to plenty of
indices appearing in the matrix elements.
The explicit form of the variational energy is too lengthy to be
presented even though $\alpha$ is fixed.

\section{Results}\label{S:R}

   First, let us concentrate our attention on the Heisenberg point
$\beta=0$.
We list in Table \ref{T:E} the stabilization energy of the
two-crackion state, ${\mit\Delta}E(L;l,0)$.
With ${\mit\Delta}E(L;l,0)<0$, a pair of crackions with their distance
$l$ may spontaneously occur in a chain of length $L$.
The crackion fluctuation is even unstable in too short chains.
Actually, at $L<8$, the VBS state itself turns out to be the
variational bound.
However, a pair of crackions come to be stabilized into a strongly
bound state with increase of $L$.
The extremely small energy stabilization at $l=1$ may be attributed
to the fact that a crackion spreads over two sites.
As $L$ increases, the crackion pairs with larger values of $l$ are
stabilized one after another.
However, the maximum distance between crackions keeping their
formation energy negative stays four at $L=1000$ and five even at
$L=10000$.
Therefore two crackions can hardly move away from each other.
Regarding $L$, as well as $l$, as a variational parameter, we
obtain the variational bound on the ground-state energy per site,
$-1.37012987$, at $L=56$ and $l=2$.
Further increase of the chain length only reduces the crackion
effect and ends in no stabilization of the ground state:
\begin{equation}
   \lim_{L\rightarrow\infty}
      \frac{E(L;l)}{L}
  =\lim_{L\rightarrow\infty}
      \frac{\langle{\rm VBS}(L)|{\cal H}|{\rm VBS}(L)\rangle}
           {L\langle{\rm VBS}(L)|{\rm VBS}(L)\rangle}
  =-  \frac{4}{3} \,.
   \label{E:limit}
\end{equation}
The ground state in the long-chain limit should be described by a
variational wave function with a finite crackion density.
Alternatively, within an approximation neglecting the interaction
between the bound crackion pairs, we may conclude that the best
density of the crackion pairs is about $1/56$.
Thus we are led to a likely ground-state picture that pairs of
crackions with $l\simeq 2$ appear in the VBS background keeping
their density about $1/56$.
The true ground state is supposed to fluctuate around this
picture.
In fact the present variational bound has not yet reached the
correct value $-1.402(1)$, which is a quantum Monte Carlo estimate
of the per-site ground-state energy at $L=56$.
The rest of the correlation may be attributed to a spatial extension
of a crackion, a breathing motion of two crackions within a pair,
and an interaction between crackion pairs.
A crackion is not equivalent to a hidden domain wall but indeed
possesses a solitonic nature \cite{Fath1}.
For example, there is a relation between a pair of hidden domain
walls and crackion pairs:
\widetext
\begin{equation}
   S_i^{\pm}S_{i+l}^{\mp}|{\rm VBS}(L)\rangle
       =-{|\Psi_{i-1,i+l-1}^{\pm\mp}(L)\rangle}
        +{|\Psi_{i-1,i+l  }^{\pm\mp}(L)\rangle}
        +{|\Psi_{i  ,i+l-1}^{\pm\mp}(L)\rangle}
        -{|\Psi_{i  ,i+l  }^{\pm\mp}(L)\rangle}\,,
   \label{E:pairflip}
\end{equation}
\narrowtext
where we assume that $l\geq 2$.
Considering that spin flips in pairs keeping the total magnetization
constant do not necessarily cause a soliton-antisoliton pair in the
hidden order (Fig. \ref{F:illust2}), the present variational
calculation is well consistent with a direct observation of the
ground state (Fig. \ref{F:snapshot}).

   Next, let us turn on the biquadratic interaction between
neighboring spins.
We plot in Fig. \ref{F:E} the stabilization energy of the
two-crackion state, ${\mit\Delta}E(L;l,\beta)$, as a function of $l$
changing $L$ and $\beta$.
The singularity observed around $L=l/2$ should be attributed to the
periodic boundary condition.
We find that the bound state of crackions, which is fully
stabilized at the Heisenberg point, becomes less stable with
increase of $\beta$.
It is a matter of course that any crackion fluctuation causes an
increase of the energy at the AKLT point $\beta=1/3$, where the VBS
state with the perfect hidden order is the exact ground state.
Interestingly, in long enough chains at $\beta>1/3$, a pair of
crackions are less stable in a bound state than when they are far
away from each other.
As $\beta$ moves away from $1/3$ toward $1$ as well as with
increase of $L$, the attractive interaction between crackions
changes into a repulsive one at small values of $l$.
Although the stabilization energy per site is more and more reduced
as $L$ increases due to the reduction of the crackion density, yet
the biquadratic interaction indeed turns the formation energy of
well-separated two crackions negative in long enough chains.
Hence a pair of crackions are set free from their bound state and are
stabilized in their wide-range breathing motion with large enough
values of $\beta$ in long enough chains, as is illustrated in Fig.
\ref{F:illust3}.
We have already stated that the strong attractive interaction between
crackions could allow the string order parameter to remain finite
against quantum fluctuations.
The collapse of the bound state of crackions and the growth of
the short-range repulsive interaction between crackions, this is a
probable mechanism of the quantum mechanical breakdown of the Haldane
phase.

\section{Summary and Discussion}\label{S:S}

   We have investigated quantum fluctuations in the ground state of
the spin-$1$ chain with the bilinear-biquadratic Hamiltonian.
With the help of quantum Monte Carlo observations of the ground
state, we have constructed a trial wave function with a pair of
crackions appearing in the VBS background.
We have demonstrated that two crackions are indeed stabilized into
a strongly bound state at the Heisenberg point.
While Monte Carlo snapshots are not available in the region of
$\beta>0$ due to the negative-sign problem, we have analytically
revealed the scenario for the quantum mechanical collapse of the
string order with $\beta$ approaching $1$.
The disappearance of the attractive interaction sets crackions free
from their bound state and the growth of a repulsive interaction
causes crackions to move away from each other.

   Besides the biquadratic interaction, there are many other
factors to cause the breakdown of the Haldane phase, such as an
alternating or anisotropic interaction
\cite{Tots1,Yama2,Bote1,Affl2,Sing1,Kato1,Yama6}.
Therefore there may be any other scenario for quantum phase
transitions.
Actually we can not describe the Haldane-to-dimer phase transition
with $\beta$ approaching $-1$ by the same scenario.
As $\beta$ moves from $1/3$ to $-1$, the bound state of crackions
remains fully stabilized.
The TB point $\beta=-1$ might be too far from the AKLT point
$\beta=1/3$ to be discussed sharing the same physics with the VBS
Hamiltonian.
However, the picture of crackion pairs moving in the VBS background
is still totally valid at the Heisenberg point \cite{Gome1,Yama3}.
Even though two crackions are stabilized into a bound state, bound
crackion pairs with a high enough density can cause the collapse
of the string order.
The breakdown mechanism of the Haldane phase with $\beta$ approaching
$-1$ may be beyond the present approximation neglecting any
interaction between crackion pairs.
Since it is hardly feasible to perform a direct calculation of numbers
of crackions keeping a definite picture of fluctuations, we expect
alternative approaches \cite{Schd1,Gome1} in the region of $\beta<0$.
Compared with the thermal breakdown, quantum phase transitions are
various and complicated.
We hope the present argument will motivate further study on the
quantum mechanical breakdown of the Haldane phase.

\acknowledgments

   The author would like to thank K. Nomura for his helpful comments.
A part of the numerical calculation was done using the facility of the
supercomputer center, Institute for Solid State Physics, University of
Tokyo.
This work is supported in part by the Japanese Ministry of Education,
Science, and Culture through the Grants-in-Aid (08740279, 09740286)
and by the Okayama Foundation for Science and Technology.

\begin{figure}
\caption{Typical fluctuations of the spin configuration.
         The black triangle denotes a domain wall in the (hidden)
         antiferromagnetic order.
         (a) A thermal fluctuation in the $S=1$ system, which allows
         the total magnetization to change and therefore leads to the
         collapse of the long-range hidden antiferromagnetic order.
         (b) A quantum fluctuation in the $S=1$ system, which keeps
         the total magnetization constant and therefore only produces
         a local effect on the hidden antiferromagnetic order.
         (c) A thermal fluctuation in the $S=1/2$ system, which is,
         in contrast with the $S=1$ case, nothing more than a local
         effect on the antiferromagnetic order.
         (d) A quantum fluctuation in the $S=1/2$ system, which
         locally disturbs the antiferromagnetic order.}
\label{F:illust1}
\end{figure}

\begin{figure}
\caption{Quantum fluctuations in the $S=1$ ground state, where the
         total magnetization is kept constant and therefore spins flip
         in pairs.
         The black triangle denotes a domain wall in the hidden
         antiferromagnetic order.
         In the case (a), a soliton-antisoliton pair appears in the
         hidden order, whereas in the case (b), there appears no
         defect in the hidden order.}
\label{F:illust2}
\end{figure}

\begin{figure}
\caption{A quantum Monte Carlo snapshot of the transformed
         two-dimensional Ising system of $L=100$ at the $S=1$
         Heisenberg point under the temperature $k_{\rm B}T=0.02$,
         where the horizontal and the vertical axes denote the chain
         and the Trotter directions corresponding to space and time,
         respectively.
         The temperature taken is low enough to represent the
         ground-state properties.
         We have set the Trotter number $n$ equal to $48$ and show
         the passage of time corresponding to $1/4k_{\rm B}T$.
         We encircle the antiphase domains which have the hidden
         antiferromagnetic order opposite to the background.
         The location of the domain walls is specified inevitably
         with some uncertainty owing to the liquid-like nature of
         the VBS state.}
\label{F:snapshot}
\end{figure}

\begin{figure}
\caption{The stabilization energy of the two-crackion state as a
         function of the crackion-crackion distance $l$ at various
         values of the chain length $L$ and the biquadratic
         interaction $\beta$,
         ${\mit\Delta}E(L;l,\beta)\equiv E(L;l,\beta)-E(L;0,\beta)$.
         (a) $L=80$, (b) $L=160$, (c) $L=240$, (d) $L=320$.}
\label{F:E}
\end{figure}

\begin{figure}
\caption{Schematic representation of the ground-state spin
         configurations at distinct regions of the biquadratic
         interaction $\beta$.
         The black triangle denotes a domain wall in the hidden
         antiferromagnetic order.
         (a) The AKLT point $\beta=1/3$. We observe a perfect hidden
         antiferromagnetic order.
         (b) $\beta<1/3$.
         The domain walls are stabilized into a bound pair.
         The hidden antiferromagnetic order is reduced but still
         persists.
         (c) $\beta>1/3$.
         The domain walls are set free from their bound state.
         The hidden antiferromagnetic order is nonlocally broken.}
\label{F:illust3}
\end{figure}

\vfill\eject
\widetext
\begin{table}
\caption{The stabilization energy of the two-crackion state as a
         function of the crackion-crackion distance $l$ and the
         chain length $L$ at the Heisenberg point,
         ${\mit\Delta}E(L;l,0)\equiv E(L;l,0)-E(L;0,0)$.}
\begin{tabular}{rrrrrrrrrrr}
$l$&
$L= 20$ & $L= 40$ & $L= 60$ & $L= 80$ & $L=100$ &
$L=120$ & $L=140$ & $L=160$ & $L=180$ & $L=200$ \\
\hline
$1$ &
${\mbox -}0.00000$ & ${\mbox -}0.00000$ & ${\mbox -}0.00000$ & ${\mbox -}0.00000$ & ${\mbox -}0.00000$ &
${\mbox -}0.00000$ & ${\mbox -}0.00000$ & ${\mbox -}0.00000$ & ${\mbox -}0.00000$ & ${\mbox -}0.00000$ \\
$2$ &
$ 0.20512$ & ${\mbox -}1.33333$ & ${\mbox -}2.20290$ & ${\mbox -}2.76190$ & ${\mbox -}3.15152$ &
${\mbox -}3.43860$ & ${\mbox -}3.65891$ & ${\mbox -}3.83333$ & ${\mbox -}3.97484$ & ${\mbox -}4.09195$ \\
$3$ &
$ 2.21638$ & $ 1.43089$ & $ 0.73563$ & $ 0.11594$ & ${\mbox -}0.43986$ &
${\mbox -}0.94118$ & ${\mbox -}1.39564$ & ${\mbox -}1.80952$ & ${\mbox -}2.18803$ & ${\mbox -}2.53552$ \\
$4$ &
$ 2.57370$ & $ 2.44126$ & $ 2.31010$ & $ 2.18079$ & $ 2.05330$ &
$ 1.92758$ & $ 1.80360$ & $ 1.68132$ & $ 1.56071$ & $ 1.44173$ \\
$5$ &
$ 2.72694$ & $ 2.71098$ & $ 2.69141$ & $ 2.67187$ & $ 2.65236$ &
$ 2.63289$ & $ 2.61345$ & $ 2.59403$ & $ 2.57465$ & $ 2.55530$ \\
$6$ &
$ 2.67604$ & $ 2.69500$ & $ 2.69236$ & $ 2.68971$ & $ 2.68706$ &
$ 2.68441$ & $ 2.68177$ & $ 2.67912$ & $ 2.67648$ & $ 2.67383$ \\
$7$ &
$ 2.60728$ & $ 2.71581$ & $ 2.71546$ & $ 2.71511$ & $ 2.71476$ &
$ 2.71441$ & $ 2.71407$ & $ 2.71372$ & $ 2.71337$ & $ 2.71302$ \\
$8$ &
$ 2.28123$ & $ 2.70902$ & $ 2.70898$ & $ 2.70893$ & $ 2.70889$ &
$ 2.70884$ & $ 2.70880$ & $ 2.70875$ & $ 2.70871$ & $ 2.70866$ \\
\hline
$l$ &
$L=220$ & $L=240$ & $L=260$ & $L=280$ & $L=300$ &
$L=320$ & $L=340$ & $L=360$ & $L=380$ & $L=400$ \\
\hline
$1$ &
${\mbox -}0.00000$ & ${\mbox -}0.00000$ & ${\mbox -}0.00000$ & ${\mbox -}0.00000$ & ${\mbox -}0.00000$ &
${\mbox -}0.00000$ & ${\mbox -}0.00000$ & ${\mbox -}0.00000$ & ${\mbox -}0.00000$ & ${\mbox -}0.00000$ \\
$2$ &
${\mbox -}4.19048$ & ${\mbox -}4.27451$ & ${\mbox -}4.34703$ & ${\mbox -}4.41026$ & ${\mbox -}4.46586$ &
${\mbox -}4.51515$ & ${\mbox -}4.55914$ & ${\mbox -}4.59864$ & ${\mbox -}4.63430$ & ${\mbox -}4.66667$ \\
$3$ &
${\mbox -}2.85564$ & ${\mbox -}3.15152$ & ${\mbox -}3.42579$ & ${\mbox -}3.68075$ & ${\mbox -}3.91837$ &
${\mbox -}4.14035$ & ${\mbox -}4.34820$ & ${\mbox -}4.54321$ & ${\mbox -}4.72655$ & ${\mbox -}4.89922$ \\
$4$ &
$ 1.32436$ & $ 1.20856$ & $ 1.09429$ & $ 0.98153$ & $ 0.87025$ &
$ 0.76042$ & $ 0.65201$ & $ 0.54499$ & $ 0.43934$ & $ 0.33503$ \\
$5$ &
$ 2.53599$ & $ 2.51670$ & $ 2.49744$ & $ 2.47822$ & $ 2.45903$ &
$ 2.43986$ & $ 2.42073$ & $ 2.40163$ & $ 2.38256$ & $ 2.36352$ \\
$6$ &
$ 2.67119$ & $ 2.66854$ & $ 2.66590$ & $ 2.66325$ & $ 2.66061$ &
$ 2.65797$ & $ 2.65533$ & $ 2.65269$ & $ 2.65005$ & $ 2.64740$ \\
$7$ &
$ 2.71267$ & $ 2.71232$ & $ 2.71197$ & $ 2.71163$ & $ 2.71128$ &
$ 2.71093$ & $ 2.71058$ & $ 2.71023$ & $ 2.70988$ & $ 2.70954$ \\
$8$ &
$ 2.70862$ & $ 2.70858$ & $ 2.70853$ & $ 2.70849$ & $ 2.70844$ &
$ 2.70840$ & $ 2.70835$ & $ 2.70831$ & $ 2.70826$ & $ 2.70822$ \\
\hline
$l$ &
$L=420$ & $L=440$ & $L=460$ & $L=480$ & $L=500$ &
$L=520$ & $L=540$ & $L=560$ & $L=580$ & $L=600$ \\
\hline
$1$ &
${\mbox -}0.00000$ & ${\mbox -}0.00000$ & ${\mbox -}0.00000$ & ${\mbox -}0.00000$ & ${\mbox -}0.00000$ &
${\mbox -}0.00000$ & ${\mbox -}0.00000$ & ${\mbox -}0.00000$ & ${\mbox -}0.00000$ & ${\mbox -}0.00000$ \\
$2$ &
${\mbox -}4.69617$ & ${\mbox -}4.72316$ & ${\mbox -}4.74797$ & ${\mbox -}4.77083$ & ${\mbox -}4.79198$ &
${\mbox -}4.81159$ & ${\mbox -}4.82984$ & ${\mbox -}4.84685$ & ${\mbox -}4.86275$ & ${\mbox -}4.87764$ \\
$3$ &
${\mbox -}5.06215$ & ${\mbox -}5.21612$ & ${\mbox -}5.36185$ & ${\mbox -}5.50000$ & ${\mbox -}5.63113$ &
${\mbox -}5.75578$ & ${\mbox -}5.87440$ & ${\mbox -}5.98742$ & ${\mbox -}6.09524$ & ${\mbox -}6.19820$ \\
$4$ &
$ 0.23203$ & $ 0.13033$ & $ 0.02989$ & ${\mbox -}0.06931$ & ${\mbox -}0.16728$ &
${\mbox -}0.26406$ & ${\mbox -}0.35966$ & ${\mbox -}0.45411$ & ${\mbox -}0.54742$ & ${\mbox -}0.63962$ \\
$5$ &
$ 2.34451$ & $ 2.32553$ & $ 2.30658$ & $ 2.28767$ & $ 2.26878$ &
$ 2.24992$ & $ 2.23109$ & $ 2.21230$ & $ 2.19353$ & $ 2.17479$ \\
$6$ &
$ 2.64477$ & $ 2.64213$ & $ 2.63949$ & $ 2.63685$ & $ 2.63421$ &
$ 2.63157$ & $ 2.62894$ & $ 2.62630$ & $ 2.62366$ & $ 2.62103$ \\
$7$ &
$ 2.70919$ & $ 2.70884$ & $ 2.70849$ & $ 2.70814$ & $ 2.70779$ &
$ 2.70745$ & $ 2.70710$ & $ 2.70675$ & $ 2.70640$ & $ 2.70605$ \\
$8$ &
$ 2.70817$ & $ 2.70813$ & $ 2.70808$ & $ 2.70804$ & $ 2.70799$ &
$ 2.70795$ & $ 2.70791$ & $ 2.70786$ & $ 2.70782$ & $ 2.70777$ \\
\hline
$l$ &
$L=620$ & $L=640$ & $L=660$ & $L=680$ & $L=700$ &
$L=720$ & $L=740$ & $L=760$ & $L=780$ & $L=800$ \\
\hline
$1$ &
${\mbox -}0.00000$ & ${\mbox -}0.00000$ & ${\mbox -}0.00000$ & ${\mbox -}0.00000$ & ${\mbox -}0.00000$ &
${\mbox -}0.00000$ & ${\mbox -}0.00000$ & ${\mbox -}0.00000$ & ${\mbox -}0.00000$ & ${\mbox -}0.00000$ \\
$2$ &
${\mbox -}4.89162$ & ${\mbox -}4.90476$ & ${\mbox -}4.91715$ & ${\mbox -}4.92884$ & ${\mbox -}4.93989$ &
${\mbox -}4.95035$ & ${\mbox -}4.96028$ & ${\mbox -}4.96970$ & ${\mbox -}4.97865$ & ${\mbox -}4.98718$ \\
$3$ &
${\mbox -}6.29662$ & ${\mbox -}6.39080$ & ${\mbox -}6.48101$ & ${\mbox -}6.56749$ & ${\mbox -}6.65047$ &
${\mbox -}6.73016$ & ${\mbox -}6.80674$ & ${\mbox -}6.88041$ & ${\mbox -}6.95131$ & ${\mbox -}7.01961$ \\
$4$ &
${\mbox -}0.73072$ & ${\mbox -}0.82075$ & ${\mbox -}0.90973$ & ${\mbox -}0.99767$ & ${\mbox -}1.08459$ &
${\mbox -}1.17051$ & ${\mbox -}1.25544$ & ${\mbox -}1.33941$ & ${\mbox -}1.42242$ & ${\mbox -}1.50450$ \\
$5$ &
$ 2.15608$ & $ 2.13740$ & $ 2.11876$ & $ 2.10014$ & $ 2.08155$ &
$ 2.06299$ & $ 2.04446$ & $ 2.02596$ & $ 2.00749$ & $ 1.98904$ \\
$6$ &
$ 2.61839$ & $ 2.61576$ & $ 2.61312$ & $ 2.61049$ & $ 2.60786$ &
$ 2.60522$ & $ 2.60259$ & $ 2.59996$ & $ 2.59733$ & $ 2.59470$ \\
$7$ &
$ 2.70570$ & $ 2.70536$ & $ 2.70501$ & $ 2.70466$ & $ 2.70431$ &
$ 2.70396$ & $ 2.70361$ & $ 2.70327$ & $ 2.70292$ & $ 2.70257$ \\
$8$ &
$ 2.70773$ & $ 2.70768$ & $ 2.70764$ & $ 2.70759$ & $ 2.70755$ &
$ 2.70750$ & $ 2.70746$ & $ 2.70741$ & $ 2.70737$ & $ 2.70733$ \\
\hline
$l$ &
$L=820$ & $L=840$ & $L=860$ & $L=880$ & $L= 900$ &
$L=920$ & $L=940$ & $L=960$ & $L=980$ & $L=1000$ \\
\hline
$1$ &
${\mbox -}0.00000$ & ${\mbox -}0.00000$ & ${\mbox -}0.00000$ & ${\mbox -}0.00000$ & ${\mbox -}0.00000$ &
${\mbox -}0.00000$ & ${\mbox -}0.00000$ & ${\mbox -}0.00000$ & ${\mbox -}0.00000$ & ${\mbox -}0.00000$ \\
$2$ &
${\mbox -}4.99531$ & ${\mbox -}5.00306$ & ${\mbox -}5.01046$ & ${\mbox -}5.01754$ & ${\mbox -}5.02432$ &
${\mbox -}5.03081$ & ${\mbox -}5.03704$ & ${\mbox -}5.04301$ & ${\mbox -}5.04875$ & ${\mbox -}5.05426$ \\
$3$ &
${\mbox -}7.08544$ & ${\mbox -}7.14894$ & ${\mbox -}7.21022$ & ${\mbox -}7.26941$ & ${\mbox -}7.32660$ &
${\mbox -}7.38190$ & ${\mbox -}7.43540$ & ${\mbox -}7.48718$ & ${\mbox -}7.53733$ & ${\mbox -}7.58592$ \\
$4$ &
${\mbox -}1.58567$ & ${\mbox -}1.66592$ & ${\mbox -}1.74529$ & ${\mbox -}1.82379$ & ${\mbox -}1.90142$ &
${\mbox -}1.97821$ & ${\mbox -}2.05417$ & ${\mbox -}2.12931$ & ${\mbox -}2.20364$ & ${\mbox -}2.27719$ \\
$5$ &
$ 1.97063$ & $ 1.95225$ & $ 1.93389$ & $ 1.91556$ & $ 1.89727$ &
$ 1.87900$ & $ 1.86076$ & $ 1.84255$ & $ 1.82436$ & $ 1.80621$ \\
$6$ &
$ 2.59207$ & $ 2.58944$ & $ 2.58681$ & $ 2.58418$ & $ 2.58155$ &
$ 2.57892$ & $ 2.57629$ & $ 2.57367$ & $ 2.57104$ & $ 2.56841$ \\
$7$ &
$ 2.70222$ & $ 2.70187$ & $ 2.70152$ & $ 2.70118$ & $ 2.70083$ &
$ 2.70048$ & $ 2.70013$ & $ 2.69978$ & $ 2.69944$ & $ 2.69909$ \\
$8$ &
$ 2.70728$ & $ 2.70724$ & $ 2.70719$ & $ 2.70715$ & $ 2.70710$ &
$ 2.70706$ & $ 2.70701$ & $ 2.70697$ & $ 2.70692$ & $ 2.70688$ \\
\end{tabular}
\label{T:E}
\end{table}

\end{document}